\documentclass[preprint]{aastex701}
\submitjournal{ApJ}
\received{2026 Apr 11}
\accepted{2026 May 14}
\usepackage{upgreek}
\newcommand{\mus}{$\upmu$s}

\begin{document}

\title{Measuring contributions from single and multiple atmospheric secondary cosmic rays in the {\it Princess Sirindhorn Neutron Monitor} using cross-counter neutron time delay distributions}

\author[orcid=0000-0002-3776-072X]{Warit Mitthumsiri}
\affiliation{Department of Physics, Faculty of Science, Mahidol University, Bangkok, Thailand}
\email{warit.mit@mahidol.ac.th}  

\author[orcid=0000-0001-7771-4341]{Alejandro S\'aiz}
\affiliation{Department of Physics, Faculty of Science, Mahidol University, Bangkok, Thailand}
\email[show]{alejandro.sai@mahidol.ac.th}
\correspondingauthor{Alejandro S\'aiz}

\author[orcid=0000-0003-3414-9666]{David Ruffolo}
\affiliation{Department of Physics, Faculty of Science, Mahidol University, Bangkok, Thailand}
\email{david.ruf@mahidol.ac.th}

\author[orcid=0000-0001-7929-810X]{Paul Evenson}
\affiliation{Bartol Research Institute, Department of Physics and Astronomy, University of Delaware, Newark, DE, USA}
\email{evenson@udel.edu}

\author[orcid=0000-0003-4865-6968]{Pierre-Simon Mangeard}
\affiliation{Bartol Research Institute, Department of Physics and Astronomy, University of Delaware, Newark, DE, USA}
\email{mangeard@udel.edu}

\author[orcid=0000-0002-1664-5845]{Waraporn Nuntiyakul}
\affiliation{Department of Physics and Materials Science, Faculty of Science, Chiang Mai University, Chiang Mai, Thailand}
\email{waraporn.n@cmu.ac.th}

\author[orcid=0000-0002-4862-2015]{Chanoknan Banglieng}
\affiliation{Division of Physics, Faculty of Science and Technology, Rajamangala University of Technology Thanyaburi, Pathum Thani, Thailand}
\email{chanoknan@rmutt.ac.th}

\begin{abstract}

Neutron monitors (NMs) are ground-based devices designed to measure cosmic-ray count rates by monitoring atmospheric neutrons from cosmic-ray showers. 
We present results from new electronics that have recorded cross-counter time delay histograms for the {\it Princess Sirindhorn Neutron Monitor} (PSNM) at the summit of Doi Inthanon, Thailand.
From these histograms, we have extracted the cross-counter leader fraction ($L$)
and corrected it for atmospheric effects.
For large counter separation, we measure nearly constant 
$L\approx0.997$, implying that 0.3\% of counts in one counter are temporally associated with later counts on a given distant counter.
Monte Carlo simulations confirm that individual secondary particles cannot account for the 
associated counts
at large counter separation, which instead requires a contribution from multiple secondary particles in the same cosmic ray shower that is apparently independent of distance over 3 to 7.5 m.
We infer that $\approx$4.5\% of PSNM counts are associated with a later count in at least one of its 18 counters from a different secondary particle in the same shower.
Monte Carlo simulations of atmospheric showers and NM yield functions can be validated using our measurements of neutron multiplicity across counters and the contributions of single and multiple secondary particles.
These measurements also improve understanding of the single-counter $L$, which has been used for precise tracking of cosmic-ray spectral variations and extending the range of NM observations to higher energies.
\end{abstract}


\section{Introduction} \label{sec:Intro}
High energy particles in space known as cosmic rays (CRs) constantly interact with the Earth's upper atmosphere, creating showers of atmospheric secondary particles that can propagate to sea level. Neutrons are a component of hadronic CR showers that can be detected by neutron monitors (NMs) \citep{Simpson2000}, typically composed of a polyethylene reflector surrounding lead rings around moderated neutron counter tubes, which are called ``counters'' here. As an atmospheric cosmic ray neutron (or sometimes a proton or other particle) interacts with a lead nucleus, some tertiary neutrons are produced and detected in the counters \citep[see e.g.,][]{Mangeard2016}. NMs are the premier tool for precisely tracking time variations in the GeV-range Galactic cosmic ray (GCR) flux, which provide remote sensing of plasma and magnetic field conditions in the solar wind near Earth and throughout the heliosphere \citep{Moraal2000}, including monitoring in real time for space weather applications \citep{Kuwabara2006b,Kuwabara2006a,Shea2012,Crosby2024}.

NMs were originally designed to measure the count rates of the tertiary neutrons, but not their energies, making spectral measurement using a single monitor station challenging.  
In some sense, 
we can use the Earth's magnetic field as a spectrometer, as it sets the ``cutoff'' (threshold) rigidity required for primary 
CRs to reach a given geographical location \citep{Smart2000},
where the rigidity $P=pc/q$ expresses the CR momentum $p$ per charge $q$.
Then a given NM count rate is related to the integral CR flux above the cutoff rigidity weighted by a yield function \citep[e.g.,][]{Clem2000,Mishev2013,Mangeard2016}.
In principle, count rates from different NMs can be compared to determine the cosmic ray spectral shape or spectral index; however, combining data from different stations involved systematic errors in  cross-calibration \citep{Aiemsa-adEA15,TimeDelay}. During the past decade, the electronics of many NM stations, including the {\it Princess Sirindhorn Neutron Monitor} (PSNM), have been upgraded with timing capabilities, enabling time-delay measurements between successive neutron detections \citep{Bieber2004,Balabin2008,Kollar2011,TimeDelay,Strauss2020}.
Because more energetic primary CRs generate more energetic secondary particles, which in turn generate a higher multiplicity of tertiary neutrons, 
analyzing the time-delay distribution of successive detections in the same counter to extract the ``leader fraction'' (equivalent to the inverse multiplicity) allowed PSNM and subsequently other NMs to continuously monitor CR spectral variations with high precision, including variations associated with solar storms (and Forbush decreases in cosmic ray intensity), 27-day variations with solar rotation, and the 11-year solar activity cycle \citep{TimeDelay,MangeardEA16b,BangliengApJ2020,Mishev2024,Muangha2024,Khamphakdee2025,Mitthumsiri2025,Hayashi2026}.

In this work, we extend the measurement technique to analyze time delays between two events detected by any pair of counters at PSNM, using upgraded electronics with absolute timing and data acquisition software that produces cross-counter time-delay histograms, from which we can define the cross-counter leader fraction $L_{ij}$ of neutron counts in counter $i$ that were not temporally associated with a following count in counter $j$, i.e., that were not associated with the same primary CR. 

The Monte Carlo simulation technique has previously been used to determine the yield function for NM count rates \citep[e.g.,][]{Clem2000,Mishev2013,Mangeard2016} and leader fractions \citep{MangeardEA16b}. Here we also simulate the interactions of individual secondary particles with the PSNM, and confirm that isolated secondary particles cannot account for the measured cross-counter multiplicity at large counter separation.
We infer that the near-counter $L_{ij}$ values are dominated by the effects of tertiary neutrons produced from the same atmospheric particle (a single secondary), while far-counter $L_{ij}$ values are dominated by multiple atmospheric particles (multiple secondaries) from the same shower created by a primary CR\@. In this way we can study the statistics of temporally associated neutron counts produced from either single or multiple CR secondaries.

\section{Detector Configuration and Data Collection} \label{sec:Dataset}

\subsection{Princess Sirindhorn Neutron Monitor Configuration}

\begin{figure}
    \centering
    \includegraphics[width=0.92\textwidth]{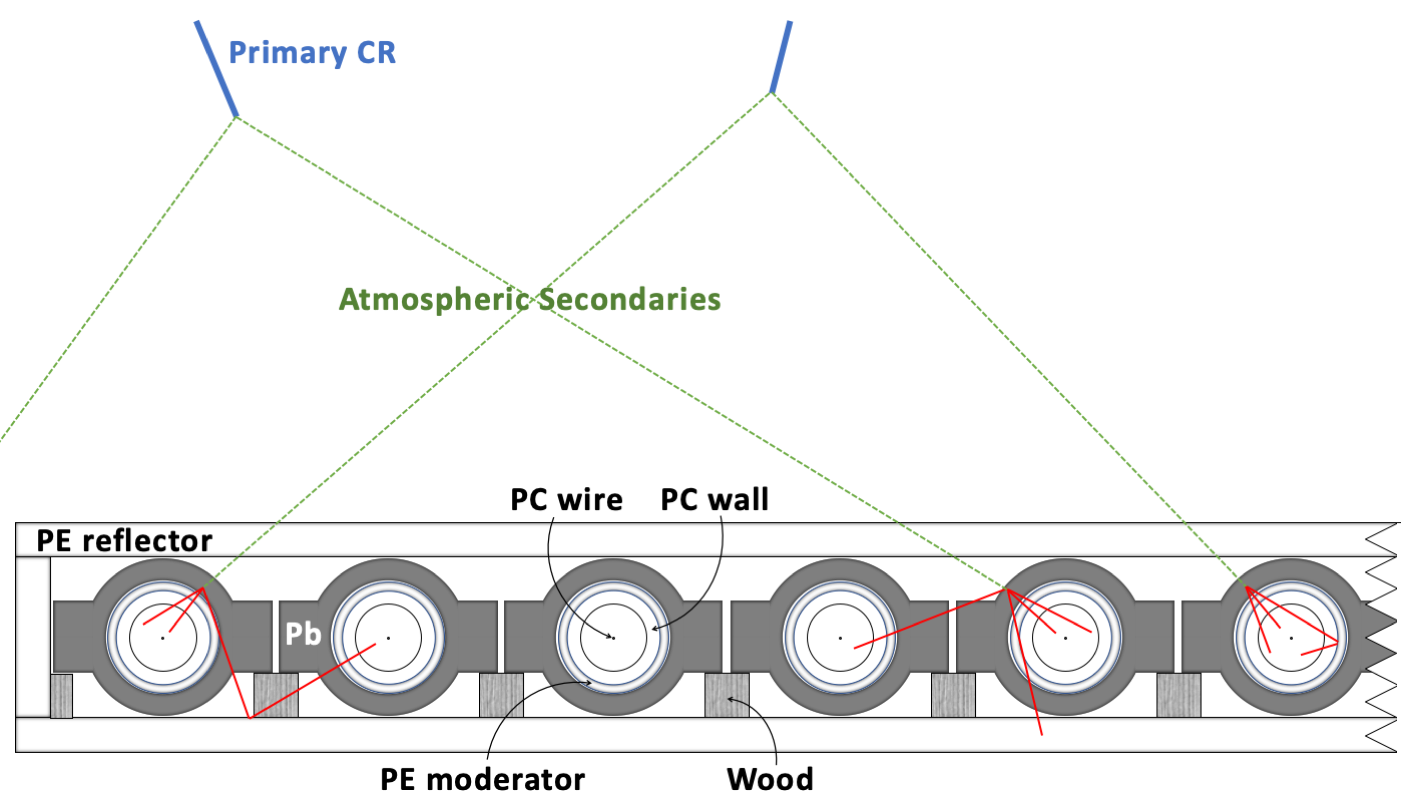}
    \caption{Schematic of ground-based neutron detection in an NM64 neutron monitor design.  
    Primary CRs (thick blue lines) interact in Earth's atmosphere, producing cascades (showers) of atmospheric secondaries (dotted green lines) which, especially for secondary neutrons, may interact with the lead (Pb) producer inside the monitor to produce multiple MeV-range tertiary neutrons (red lines) that can be moderated by polyethylene (PE) and detected by one or more of the proportional counters (PC\@).  
    We refer to neutron detection in different counters (labeled $i$ and $j$) associated with the same primary CR (i.e., temporally associated) as cross-counter multiplicity, the inverse of which is the leader fraction, $L_{ij}$.
    Using $L_{ij}$ from counters at various separations $\Delta\equiv|i-j|$, we can statistically distinguish the contributions of single and multiple atmospheric secondaries to the neutron multiplicity.}
    \label{fig:orient}
\end{figure}

PSNM is located at the summit of Doi Inthanon, Thailand, at an altitude of 2560 m and geographical coordinates 18.59$^\circ$ N, 98.49$^\circ$ E. It is an ``18NM64'' with 18 BP-28 neutron counters (proportional counters filled with $^{10}$BF$_3$ gas) in the standard NM64 configuration, arranged horizontally in a single row with no intervening polyethylene. 
The separation between adjacent counters is 50~cm.
PSNM's location at high rigidity cutoff ($\approx$17~GV) helps filter out the much greater and fluctuating flux of low-energy CRs, and its simple arrangement makes PSNM a good station for the first cross-counter time-delay analysis. In this arrangement, there are 16 middle and 2 end counters, denoted as ``$m$'' and ``$e$'', respectively. PSNM's cross-sectional components are illustrated in Figure~\ref{fig:orient}, showing an example of 1 ``$e$'' and 5 ``$m$'' counters surrounded by lead rings labeled as Pb. 

Figure~\ref{fig:orient} also indicates that the same atmospheric secondary (green line) can interact with a lead ring to produce a group of tertiary neutrons (red lines) which are scattered and detected by different nearby counters. Since the $m$ counters are surrounded by more lead producer, having nearby lead on both sides while the $e$ counters have less lead on one side, the $m$ counters detect higher count rates than the $e$ counters by 16\% \citep[i.e., $C^e/C^m\approx0.86$;][]{Mangeard2016}. According to our simulations, tertiary neutrons normally do not diffuse to a distance greater than $\sim$3~m, or $\sim$6 times the counter separation distance.  
In Section~\ref{sec:Results} we will provide evidence that the statistical temporal association of events detected in a pair of counters which are significantly more than 6 counters apart can only be explained by the detection of two different atmospheric secondaries produced by the same primary CR.

\subsection{Absolute Timing of Neutron Counts and Cross-Counter Time-Delay Histograms}

\begin{figure}
    \centering
    \includegraphics[width=0.9\textwidth]{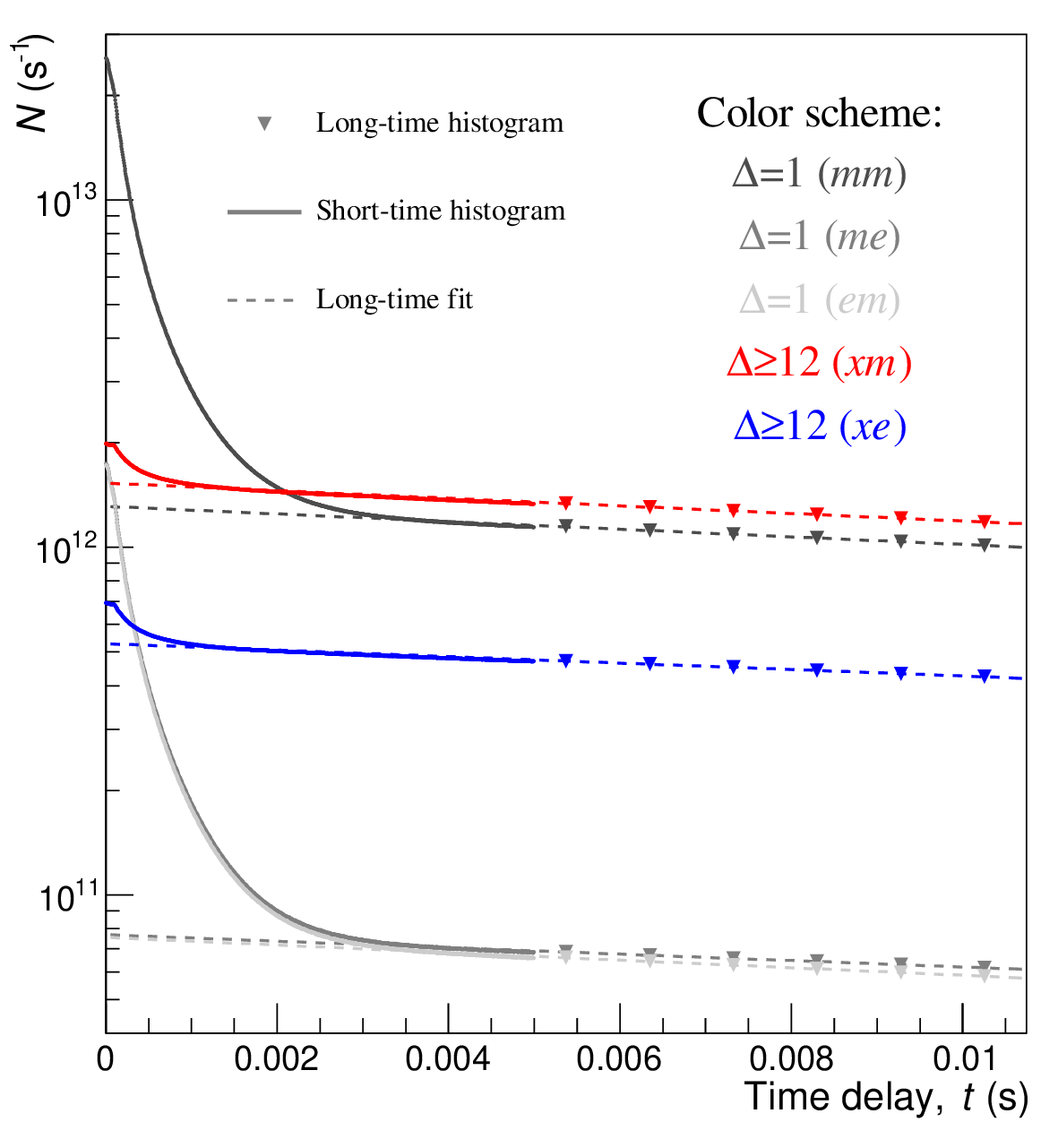}
    \caption{
    Cross-counter neutron time-delay distributions from histograms accumulated over $\approx$6 years at PSNM\@. We show examples for counter separations $\Delta=1$ (adjacent counters) and $\Delta\geq12$ (far counters).  
    Data are not normalized for the number of counter pairs (see text).
    Here $t$ is the time delay from a neutron detection in counter $i$ to the next detection in counter $j$. 
    The two-letter notation indicates the counter positions, e.g., $mm$ means that $i$ and $j$ are both middle counters, $me$ means that $i$ is a middle counter while $j$ is an end counter, $xm$ includes any counter $i$ while $j$ is a middle counter, etc.  
    At long $t$, these distributions are well described by exponential functions (dashed lines), corresponding to physically unrelated chance coincidences.
    However, the enhancement above the exponential at short time delays is due to multiple 
    neutron counts from the same primary CR\@.
    We will provide evidence that the strong relative enhancement for $\Delta=1$ is dominated by interactions from a single atmospheric secondary CR, while the weaker relative enhancement for $\Delta\geq12$ indicates the detection of multiple secondaries in the NM (see Figure \ref{fig:orient}).}
    \label{fig:xtd2}
\end{figure}

A major electronics upgrade in 2015 has allowed PSNM to measure the absolute timing of event detection in all neutron counters, referenced to the GPS time \citep{Saiz2017}.  The timing accuracy is approximately $\pm$3~\mus.
For each count in each counter $i$, we examine the subsequent data in each counter $j$, recording the time delay $t$ until a count is recorded in $j$ by incrementing the ``time-delay'' histograms.
Each time delay is recorded in both a short-time histogram (with 1024 bins from $t=0$ to $\approx$5.0~ms) and a long-time histogram (with 1024 bins from $t=0$ to $\approx$1.0~s). 
Histograms are saved to files and reset hourly.

At the time of the electronics upgrade in late 2015, to reduce the number of recorded histograms, time-delay data from all pairs of counters with the same separation $\Delta\equiv|i-j|$ were grouped together, so 18 sets of cross-counter time-delay histograms were recorded hourly for $\Delta=0$ to 17. For example, $\Delta=1$ includes all pairs of adjacent counters, while $\Delta=17$ only includes the two counters at each end of the row [i.e., only $(i,j)=(1,18)$ and (18,1)]. We then recognized that time-delay histogram data depend on whether the $i$ and $j$ counters are middle ($m$) or end ($e$) counters, so since August 2017 we have collected 50 sets of hourly histograms, each of which has a unique combination of $\Delta$ and counter positions ($m$ or $e$), indicated by a two-letter notation referring to the first and second counter, respectively.  For example, for adjacent counters (i.e., $\Delta=1$), the $mm$, $me$, and $em$ configurations are possible, but $ee$ is not. 
Software and hardware changes sometimes interrupted our data collection, such as problems with absolute timing between September and December 2018 and during the first half of 2021, and reverting back to 18 histograms during March to June 2019. 
Here we show our valid data between April 2018 and December 2024.

We combine such histogram data from all hours (after cleaning discussed in Section~\ref{sec:Analysis}) as shown in Figure~\ref{fig:xtd2}, including short-time histogram data (curves) up to 5.0~ms and long-time data (symbols) above 5.0~ms, normalized by the bin width.  The figure shows data for example cases of
$\Delta=1$ (adjacent counters) and $\Delta\ge12$ (combining all histograms for separations of 12 or more counters). 
Here $x$ includes both $m$ and $e$ counters; for example, the combination $\Delta\geq12(xm)$ 
includes time delays for any pair of counters with the later count in a middle counter.
The absolute number of counts $N(t)$ in Figure 2 reflects the number of counter pairs.
For example, $N(t)$ is about 15 times larger for $\Delta=1 (mm)$ than for $\Delta=1 (me)$ or $\Delta=1 (em)$. 
Because the integral of $N(t)$ is the total number of counts from the first counter, and $m$ counters collect about 16\% more counts than $e$ counters as noted earlier, $me$ has a slightly higher $N(t)$ than $em$.  
In the remainder of the analysis, we will focus more on the shapes of the time-delay distributions than their amplitudes. 

For a same-counter analysis, the time-delay histograms at low $t$ are suppressed due to an electronic dead time  \citep[the inability to record a following event arriving closely in time after an initial event; see][]{TimeDelay}.
For our electronics, the dead time for the time-delay circuitry \citep[$t_d=72$ to 83~\mus;][]{BangliengApJ2020} is distinct from that of the counting circuitry \citep[19 to 29~\mus;][]{Mangeard2016}, so in principle one could distinguish ``pulses'' in the time-delay analysis as a subset of the counts.
However, in this cross-counter analysis (for $\Delta>0$) the following count is registered by a different counter from the initial count, so the electronic dead time has less effect on the time-delay histograms, and here we will not distinguish pulses from counts.

\section{\label{sec:Analysis}Cross-Counter Leader Fraction Analysis}

\subsection{Definition of Cross-Counter Leader Fractions}

In analogy with the single-counter leader fraction \citep{TimeDelay}, and based on the recording technique explained in the previous section, we define the cross-counter leader fraction $L_{ij}$ as the fraction of counts in counter $i$ that are not temporally associated (i.e., from the same cosmic ray shower) with a later count in counter $j$. We can measure $L_{ij}$ using time-delay histograms such as shown in Figure~\ref{fig:xtd2}. Above $t\approx3$~ms, the time delay between two successive counts is dominated by chance coincidence of two unassociated events from different primary CRs, for which the distribution can be characterized by an exponential functional form:
\begin{equation}
    n^{kl}_{\Delta}(t) = \beta^{kl}_{\Delta}\exp(-\alpha^{kl}_{\Delta}t).
    \label{eq:expo}
\end{equation}

The subscript $\Delta=|i-j|=0, 1, 2, ..., 17$ labels counter separation, and the superscript $kl = mm$, $me$, $em$, or $ee$ denotes the position (middle or end) of the first and second counter, respectively;
we do not accumulate $18^2=324$ histograms for every possible combination $ij$, but rather only 50 histograms for each possible combination of $\Delta$ and $kl$.
To safely avoid the times $t<3$ ms that may have a significant excess above the exponential trend,
we fit the long time-delay histograms at $t=5-200$~ms with Equation~(\ref{eq:expo}), using $\alpha^{kl}_{\Delta}$ and $\beta^{kl}_{\Delta}$ as fitting parameters. Fitting results are indicated by dashed lines in Figure~\ref{fig:xtd2}. The extrapolations of the exponential functions down to $t=0$~ms show that the measurements at $t\lesssim3$~ms deviate from the exponential trends due to the contribution from temporally associated events. Here we define the cross-counter leader fraction as:
\begin{equation}
L^{kl}_{\Delta}=\frac{\text{Number of counts in exponential}}{\text{Total number of counts}}=\frac{\int_{0}^{\infty}n^{kl}_{\Delta}(t)dt}{\Sigma_{z=1}^{z(t_h)}{(b^{kl}_{\Delta})}_{z}\Delta t_{z}+\int_{t_h}^{\infty}n^{kl}_{\Delta}(t)dt},
\label{eq:L}
\end{equation}
where $({b^{kl}_{\Delta}})_z$ is the $z^{th}$ normalized bin content of the long time-delay histogram with counter separation $\Delta$ and counter position $kl$, $\Delta t_z$ is the $z^{th}$ bin width, $t_h=200$~ms is the switching time from using the histogram contents to using fitted parameters, and $z(t_h)$ is the bin number corresponding to $t_h$. 

The term ``leader'' is carried over from the single-counter leader fraction.  
In that case, the remote electronics on each counter recorded the time delay of each neutron count relative to the previous count on that counter.
Thus the time delay $t$ referred to a time in the past, and the leader fraction $L$ (with a definition similar to Equation~(\ref{eq:L})) represented the fraction of counts that did not follow another count from the same shower, i.e., ``leaders.''  
In the same vein, $1-L$ was interpreted the fraction of ``followers'' that do follow another count from the same shower.
Previous work has stressed the leader rather than follower fraction because it is directly obtained from the exponential fit parameters, varies in the same sense as the cosmic ray spectral index \citep[e.g.,][]{Muangha2024}, and represents the inverse of the neutron multiplicity.
In the present work, there is a time reversal in the sense that the time delay as recorded by the data acquisition software refers to the time {\it after} a count on counter $i$ when the next count occurred on counter $j$, so
it might be more accurate to refer to $L_\Delta^{kl}$ as the fraction of ``trailers'' that are not followed by a temporally associated count on the second counter; however, we can view that
there is an approximate time reversal symmetry to the statistics of time delays before or after a count, and thus we will continue to refer to $L_\Delta^{kl}$ as a leader fraction and
$1-L_\Delta^{kl}$ as a follower fraction.
We can interpret $1/L_{\Delta}^{kl}$ as a cross-counter multiplicity in the sense that a value of, say, 1.1 implies that for every count on $k$ that was not followed by an associated count on $l$, there were 1.1 counts on $k$ in total.

The slope of the dashed lines in Figure~\ref{fig:xtd2} is $\alpha^{kl}_{\Delta}$, which depends strongly on the position of the second counter. 
This is the key reason for separating time-delay histograms for different counter positions ($e$ or $m$), so as to avoid combining data with different exponential trends.
Physically, $\alpha$ represents the leader count rate in the second counter (not corrected for pressure variation).
The fitted values of $\alpha$ for each hour vary slightly, mainly due to the variation of the atmospheric pressure. To combine hourly histograms into a long-time histogram such as shown in Figure~\ref{fig:xtd2}, we calculate the median of the distribution of $\alpha$ and use hours when $\alpha$ is within 1\% of the median value, which includes about half of the data. This clean-up process ensures that we combine histograms with the same exponents so that the final histogram stays in 
the exponential form at large $t$.

We define pairs of detectors with $\Delta\ge12$ as ``far counters'' and for better statistics we sometimes combine time-delay histograms for any pair of counters with a separation of at least 12, because it will be seen in Section~\ref{sec:LvsDelta} that at $\Delta\ge12$,  $L_{\Delta}$ is approximately independent of $\Delta$. Note that this far-counter combination does not occur at the data collection level, but rather by stacking the corresponding time-delay histograms. We observe that the position of the second counter has an effect on the value of the far-counter leader fraction (the proposed explanation is in Section~\ref{sec:LvsDelta}). Thus, we combine the far-counter histograms into 2 cases, $xm$ and $xe$, for which the second counter is a middle and end counter, respectively, regardless of the first counter position. Time-delay histograms of $\Delta\ge12$ for $xm$ and $xe$ are emphasized in red and blue, respectively, in Figure~\ref{fig:xtd2}.

\begin{figure}
    \centering
    \includegraphics[width=\textwidth]{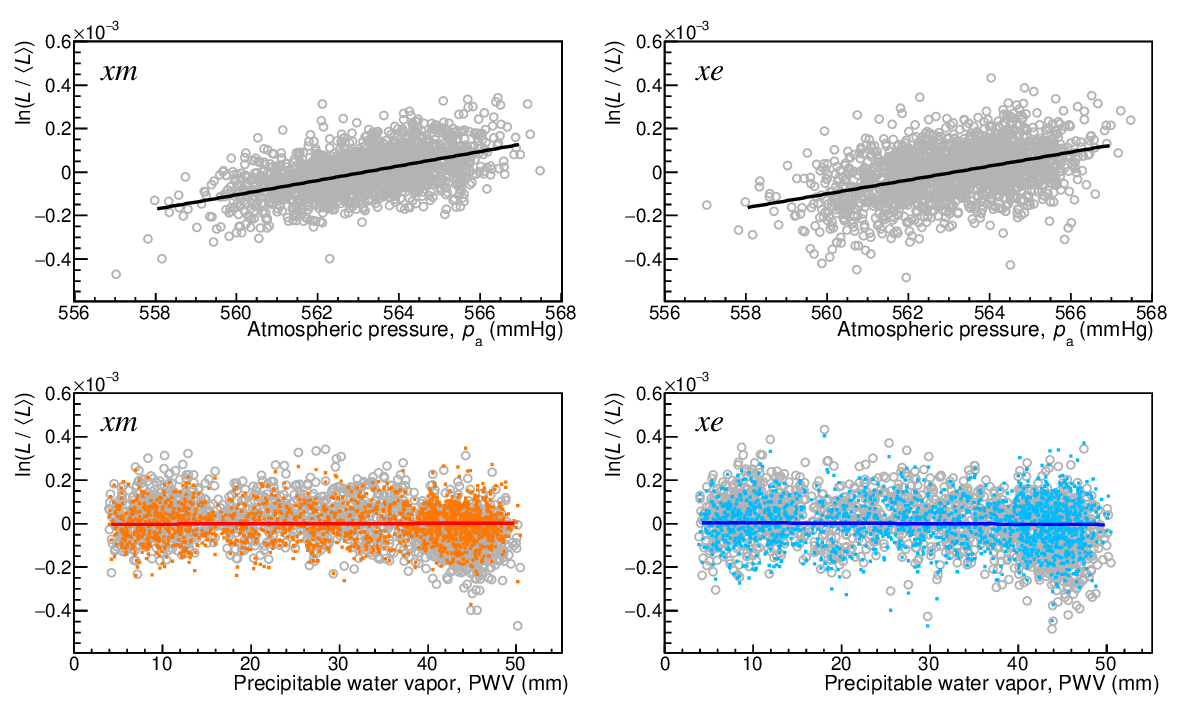}
    \caption{
    (Top panels) Scatter plots showing logarithmic variation of the far-counter leader fraction $\ln(L/\langle L\rangle)$ vs.\ atmospheric pressure $p_\mathrm{a}$ for $\Delta$$\ge$$12$, where $L$ is the daily average value and $\langle L\rangle$ is the average value of the whole analysis time period. The fitted slopes of a linear function (solid black lines) are $(3.32\pm0.13)\times10^{-5}$ mmHg$^{-1}$ for $xm$ and $(3.20\pm0.16)\times10^{-5}$ mmHg$^{-1}$ for $xe$. (Bottom panels) Similar plots vs.\ precipitable water vapor (PWV), where the grey (colored) symbols indicate values before (after) the $p_\mathrm{a}$ corrections based on the above slopes. The fitted slopes of a linear function to the corrected data (red solid lines) are $(0.90\pm1.5)\times10^{-7}$ mm$^{-1}$ for $xm$ and $(-0.21\pm1.9)\times10^{-7}$ mm$^{-1}$ for $xe$, which are consistent with zero, indicating that the pressure correction also removes the dependence on atmospheric water vapor.}
    \label{fig:XLFcorr}
\end{figure}

\subsection{Corrections for Atmospheric Pressure and Water Vapor \label{sec:corr}}

Previous work on the same-counter leader fraction has shown that there are effects of variation in atmospheric pressure and, in a tropical location, water vapor \citep{TimeDelay,Muangha2024}.  PSNM is equipped with air pressure sensors, typically recording the atmospheric pressure ($p_\mathrm{a}$) at the summit between 555--570~mmHg. We derive the precipitable water vapor (PWV) from the Global Data Assimilation System (GDAS) database and apply $\pm5$-day triangular smoothing using the same method as described in \citet{BangliengApJ2020}. 
That work also found that $p_\mathrm{a}$ and PWV exhibit strong seasonal variations at Doi Inthanon, 
and are strongly anticorrelated with one another (see also Section~\ref{sec:timeseries}). 
Here we examine these atmospheric effects (only) for the far-counter leader fraction $L=L_{\Delta\ge12}$ by observing the scatter plots of $\ln(L/\langle L\rangle)$ with $p_\mathrm{a}$ and PWV, where $L$ in this plot is the daily average value and $\langle L\rangle$ is the average value of $L$ for the whole analysis time period (Figure~\ref{fig:XLFcorr}). 

We observe clear positive correlations of $L$ with $p_\mathrm{a}$.  To quantify the relationship, we
fit the scatter plot of $\ln(L/\langle L\rangle)$ vs.\ $p_\mathrm{a}$ with a linear function, $f(p_\mathrm{a})=A(p_\mathrm{a}-p_\mathrm{ref})$
(solid black lines in the top two panels of Figure~\ref{fig:XLFcorr}) to obtain the best-fit parameters $A_0$ and $p_\mathrm{ref,0}$. The best-fit slopes $A_0$ for $xm$ and $xe$ are indicated in the Figure~\ref{fig:XLFcorr} caption. 
Then we can correct for pressure variation using
\begin{equation}\label{eq:Lcorr}    
    L_{\rm corrected}=L_{\rm uncorrected}/\exp\left[A_0(p_\mathrm{a}-p_\mathrm{ref,0})\right].
\end{equation}

There are also negative correlations of $L$ with PWV, as shown in the two bottom panels of Figure~\ref{fig:XLFcorr}. Before the $p_\mathrm{a}$ correction (grey circles), $\ln(L/\langle L\rangle)$ weakly anticorrelates with PWV\@. However, after the $p_\mathrm{a}$ correction (colored symbols) the dependence on PWV has effectively been removed, and now the fitted slopes (red solid lines) are consistent with zero, i.e., the effects of PWV can be neglected after the $p_\mathrm{a}$ correction.

\section{\label{sec:Results}Results and Discussion}

\subsection{\label{sec:LvsDelta}Cross-Counter Leader Fraction vs. Counter Separation}

\begin{figure}
    \centering
    \includegraphics[width=0.92\textwidth]{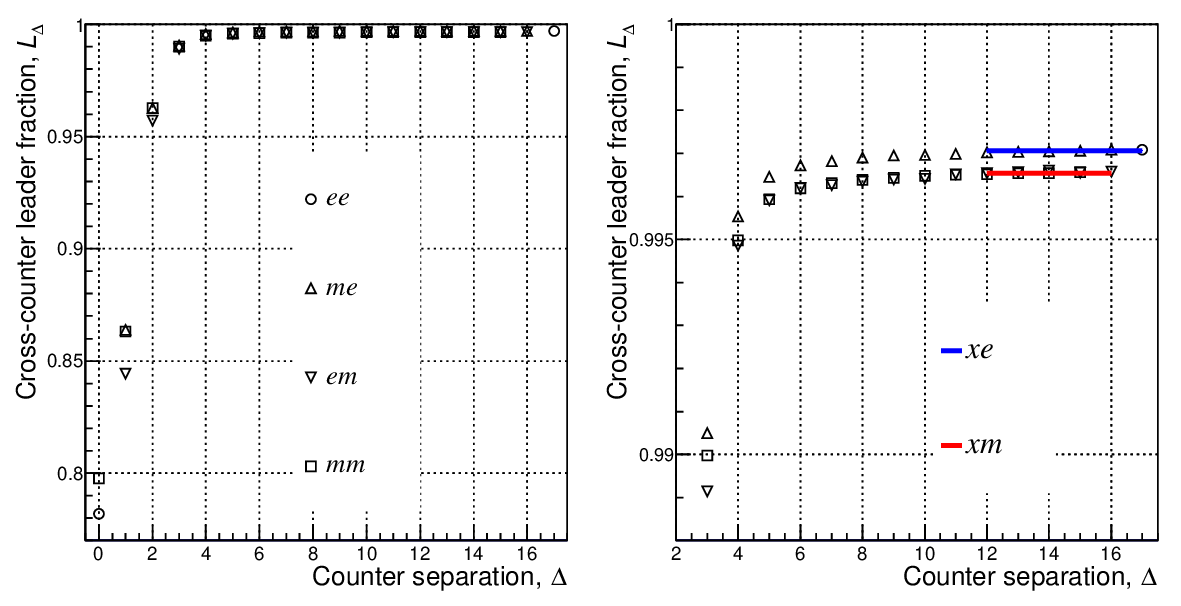}
    \caption{(Left) Cross-counter leader fraction ($L_{\Delta}$) as a function of counter separation ($\Delta$) for different pair configurations, based on valid data from April 2018 to December 2024. (Right) Zoom-in of the left panel, showing that $L_{\Delta}$ approaches a constant when $\Delta\ge12$. Here $xm$ and $xe$ denote any pair configuration for which the second counter is at the middle or at the end, respectively.}
    \label{fig:xtd1}
\end{figure}

The cross-counter leader fraction  is plotted vs.\ counter separation in Figure~\ref{fig:xtd1}, based on histograms such as those shown in Figure~\ref{fig:xtd2}.  The increasing trend implies 
that two successive counts in a pair of counters are less likely to be temporally associated (i.e., associated with the same shower) at larger separation. 
At small separation, the leader fraction value increases from $\approx$0.8 at $\Delta=0$ to $\approx$0.990 at $\Delta=3$, and finally saturates at $\approx$0.997 for $\Delta\geq12$, corresponding to a counter-to-counter distance $\geq6$~m. 
A value of $L = 1$ would correspond to totally uncorrelated neutron detections between
counters (a Poisson process). However, this is never observed, even at $\Delta\geq12$.
Neutrons from interactions of a single secondary shower particle, typically in the lead producer, are unable to diffuse over such a great distance in sufficient numbers (see also Section~\ref{sec:MC}),
so in the far-counter regime $L<1$ indicates the effect of neutron counts from different secondary particles in the same cosmic ray shower.

For small separation the precise value of $L$ depends on  whether the {\it first} counter is a middle ($m$) or an end ($e$) counter. For example, the left panel of Figure~\ref{fig:xtd1} shows that $L_{\Delta=1}^{me}\approx L_{\Delta=1}^{mm}>L_{\Delta=1}^{em}$, i.e., the first counter being $m$ results in higher value of $L$. However, for larger separation ($\Delta\ge4$), as shown on the right panel of Figure~\ref{fig:xtd1}, $L$ instead depends on the position of the {\it second} counter in the combination, i.e., $L_{\Delta}^{xe}>L_{\Delta}^{xm}$.

We can explain the dependence of the cross-counter leader fraction on the counter position (end or middle) for the small and large separations $\Delta$ as follows:
Let $C^m$ and $C^e$ be the count rates of middle and end counters, respectively, recalling that $C^m$ is greater by 16\% because a middle tube has lead producer around neighboring counters on both sides. Then let $N^{kl}_\Delta$ be the rate of counts in counter $l$ that were temporally associated with a previous count in counter $k$ at separation $\Delta$. The leader fraction can be expressed as
\begin{equation}
    L^{kl}_\Delta = 1-N^{kl}_\Delta/C^k = 1-P^{kl}_\Delta.
    \label{eq:LN}
\end{equation}

For small separation ($\Delta\le2$), two successive counts are usually due to a single atmospheric secondary particle interacting with the lead producer to create multiple tertiary neutrons.  (Evidence for this assertion will be provided in the following subsections.) 
Oscilloscope measurements with PSNM counter tubes \citep{Chaiwongkhot2024} indicate that this process is approximately independent of the position of a given pair of counters 
due to the short diffusion length of most low-energy tertiary neutrons, so we expect that all $N^{kl}_\Delta$ are very similar for $\Delta=1$ or 2.
Since $C^m>C^e$, then according to Equation~(\ref{eq:LN}), we have $L^{mx}_{\Delta}>L^{ex}_{\Delta}$.  
Indeed, for these cases we do find $(1-L^{mx}_\Delta)/(1-L^{ex}_\Delta)\approx C^e/C^m = 0.86$, as expected from this argument.

For large separations with $\Delta\ge12$, temporally associated counts are 
attributed to multiple secondary particles from the same primary cosmic ray. 
Thus $P^{kl}$ is the probability of neutron detection at $l$ from an unrelated secondary, and is simply proportional to the counting efficiency of counter $l$, i.e., its overall count rate $C^l$. This can explain why $L^{kl}_{\Delta\ge12}$ depends only on $l$, the position of the second counter. 
Quantitatively, from the right panel of Figure~\ref{fig:xtd1}, $L_{\Delta\ge12}^{xe}\approx0.99706$ and $L_{\Delta\ge12}^{xm}\approx0.99654$. According to Equation~(\ref{eq:LN}), $P^{xe}_{\Delta\ge12}/P^{xm}_{\Delta\ge12}=(1-L_{\Delta\ge12}^{xe})/(1-L_{\Delta\ge12}^{xm})=0.85$, 
which is close to the end-to-middle count rate ratio
$C^e/C^m=0.86$,
experimentally validating our interpretation.  
By extension, this argument based on a contribution of counts from different secondaries can explain why $L^{xe}_\Delta>L^{xm}_\Delta$ for all $\Delta\geq4$, with little dependence on the position of the first counter.

\begin{figure}
    \centering
    \includegraphics[width=\textwidth]{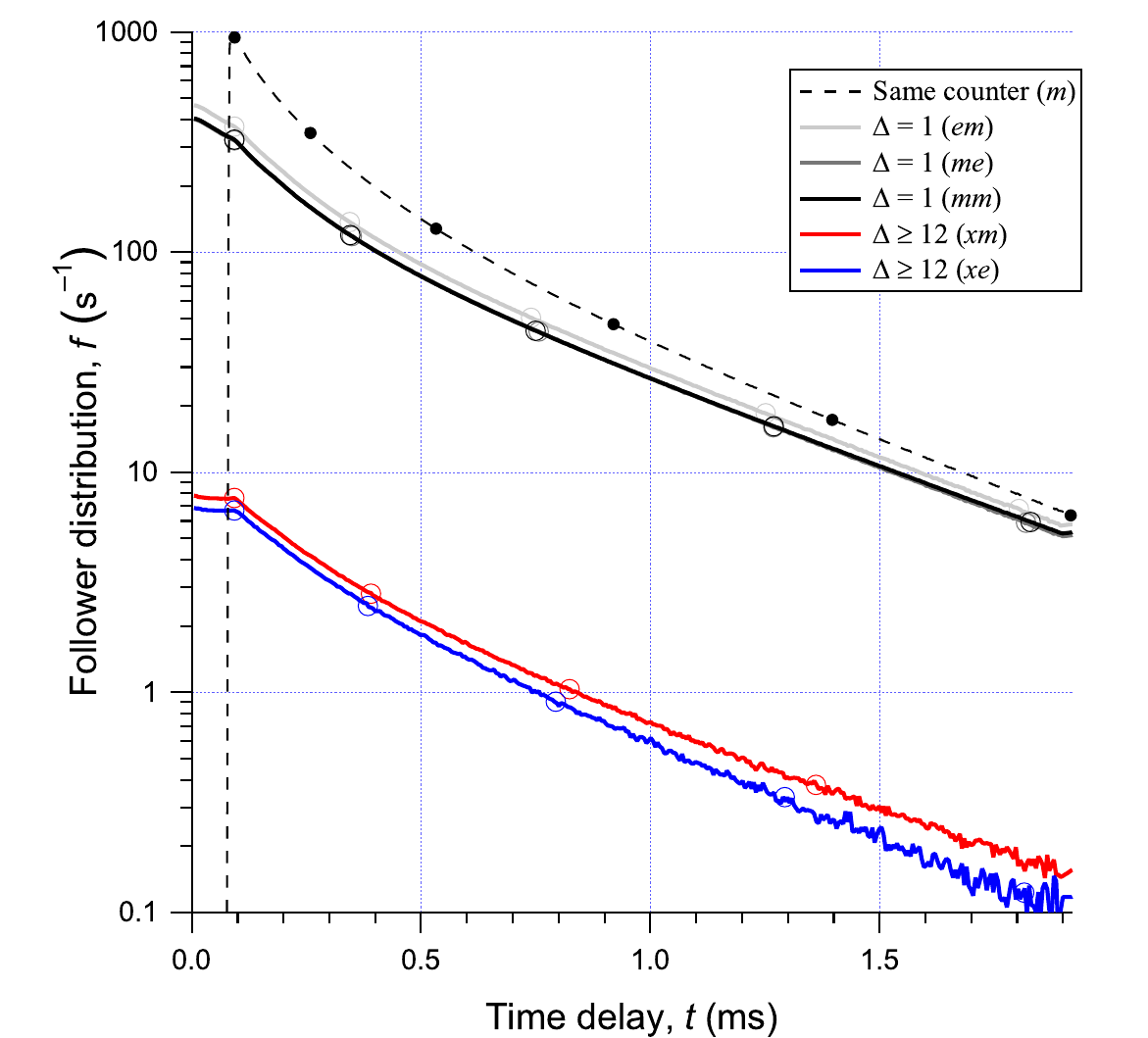}
    \caption{Follower time-delay distribution, $f(t)$, for five cross-counter combinations with counter separations $\Delta=1$ and $\Delta\geq12$ (see text for details).  The trace from Figure 4(b) of \protect\citet{TimeDelay} for same-counter $f(t)$ (dashed line) is included for comparison.  Each circle marks the value of $t$ where $f(t)$ has a value $1/\mathrm{e}$ times that at the previous circle, starting from a common reference value of $t=93$~\mus.
    Note that the traces are in the same order as the legend, and the trace for $\Delta=1\,(me)$ is overplotted by the trace for $\Delta=1\,(mm)$.
   Cross-counter time delays at small $\Delta$ are from the same secondary particle and are noticeably longer than same-counter time delays, which indicates a longer neutron propagation time inside the NM\@.  The far-counter time delays at large $\Delta$ are dominated by multiple secondaries and the distribution decays even more slowly, 
   which indicates a time delay between secondaries in some cases, 
  e.g., when the later secondary is a low-energy neutron.}
    \label{fig:fhist}
\end{figure}

\subsection{\label{sec:f(t)}Follower Time-Delay Distributions}

The follower time-delay distribution, $f(t)$, can be constructed from each of the time-delay histograms as explained by \citet{TimeDelay}. Followers can be interpreted as time-associated neutron counts due to the same cosmic ray shower, after statistically removing the effect of chance coincidences. Figure~\ref{fig:fhist} shows  examples of $f(t)$ distributions from the five histograms accumulated over $\approx $6~years as shown in Figure~\ref{fig:xtd2}. For comparison, we also plot the distribution for same-counter time delays averaged over 3~years for 8 middle counters at PSNM, taken from \citet{TimeDelay}.
The overall amplitudes differ because the total integral of each follower time-delay distribution is the ``follower fraction,'' 
$1-L^{kl}_{\Delta}$.
The effect of the time-delay dead time $t_d$ can be seen at low $t$, especially for the same-counter follower distribution, in which case no time delays can be recorded during the dead time, so the distribution abruptly starts at that time.
For the cross-counter distributions, relative time delays across different counters $i$ and $j$ can occur with $t<t_d$.  
Nevertheless, the dead time still has a minor effect
because some cosmic ray showers produce numerous counts in rapid succession on the same counter.
For example, in the case that a count on counter $i$ occurs immediately after a count on counter $j$,
then the electronics are unable to register a new count on counter $j$ for a time $t_d$.  
There is some reduction in $f(t)$ at $t<93$~\mus\ due to this effect.

To visualize the rate of decrease of a given distribution in Figure~\ref{fig:fhist}, each circle marks an $\mathrm{e}$-folding, i.e., the time delay $t$ where $f(t)$ has a value $1/\mathrm{e}$ times that at the previous circle, starting from a common reference time of $t=93$~\mus.
We can clearly see that the same-counter distribution (dashed curve) decays most rapidly.  
This distribution, between successive neutron counts in the same counter, is dominated by multiple tertiary neutrons from the interactions of a single secondary, most commonly in the lead rings surrounding that counter.  
Thus the distribution is dominated by the difference in propagation times of different neutrons from the same interaction.
This propagation time was directly measured by \citet{Chaiwongkhot2021}, in an experiment where a charged-particle detector was placed on top of PSNM counter tube 1.  
For the case of neutrons generated by a charged secondary particle, the neutron propagation time was measured relative to the arrival time of the secondary particle, and was found to be consistent with the analytic solution to a two-dimensional diffusion-absorption equation.
The same-counter follower time-delay distribution measured here decays at a similar rate to the propagation time distribution reported by \citet{Chaiwongkhot2021}.

Next, we find that the cross-counter follower distributions decay more slowly than the same-counter distribution.  
For the case of adjacent counters, $\Delta=1$, the distribution should still be dominated by detection of two tertiary neutrons from an interaction of the same secondary particle; however, in this case one of the neutrons had to travel from the lead surrounding one counter to an adjacent counter, which is typically a slightly longer distance.
This can explain why the distribution spreads to longer differences in propagation time, i.e., larger time delays $t$.  
Indeed, a measurement of this type was also performed using charged-particle detectors at PSNM, using directional coincidence measurements indicating the location where a charged secondary entered the NM (near tube 10 or 11).
The measurements are consistent with a longer propagation time for neutrons detected by a counter adjacent to the counter tube where the charged secondary entered, compared with detection in the same counter \citep{Amratisha2025}.

Finally, the far-counter follower  distributions in Figure~\ref{fig:fhist} evolve even more slowly.
These are distributions of time delays $t$ between detection of tertiary neutrons from two different secondary particles from the same cosmic ray shower.
The relatively slower decay of the distribution indicates that in some cases there is a relative delay between the two secondary particles.

While charged secondary particles arrive at ground level in a nearly planar front, close to the speed of light with a distribution in arrival times that is mostly less than  1~\mus, secondary neutrons are well known to commonly have later arrival times, especially for lower-energy neutrons \citep{Tongiorgi1949,Linsley1984}.  Specifically, previous simulations using the FLUKA code indicate that secondary neutrons that are delayed by several hundred \mus\ or more are almost exclusively at energies $<1$~MeV \citep{Schimassek2024}.
While simulations indicate that most counts at PSNM are contributed by more energetic secondary neutrons or protons \citep{Aiemsa-adEA15}, there is a small (1.5\%) contribution from secondary neutrons of energy $<10$~MeV (see Table~3 of that work).
Thus it seems likely that the slower rate of decline in follower distributions for large counter separation can be attributed to multiple secondaries, with occasional cases of a relatively long delay before the arrival of another secondary that is a low-energy neutron.

\subsection{\label{sec:MC}Monte Carlo Estimation of Cross-Counter Multiplicity from a Single Secondary}

Using the Monte Carlo (MC) particle transport simulation package FLUKA 4 \citep{BATTISTONI201510,AhdidaEA2022}, we estimated the cross-counter multiplicity produced by single secondary particles impacting the 18NM64 located at Doi Inthanon. 
As the main contribution of a NM count rate comes from neutrons in the energy range 10~MeV---1~GeV (estimated as $\approx 63\%$ at PSNM \citep{Aiemsa-adEA15}), we simulated the NM cross-multiplicity for individual mono-energetic vertical neutrons of 10, 50, 100 and 300 MeV, and 1 GeV\@.
No significant contribution to cross-multiplicity with $\Delta>6$ was observed. 
Moreover, considering that the multiplicity in a NM increases with the energy of the incoming secondary particles, and that nucleons with an energy above 1 GeV contribute $\approx 31\%$ of the count rate at PSNM~\citep{Aiemsa-adEA15}, we also estimated the cross-multiplicity produced by incoming nucleons in the energy range 1--500~GeV\@.
As protons and neutrons with an energy above 1~GeV produce similar response in a NM \citep[and references therein]{MangeardEA16b}, only interactions from incoming secondary neutrons were simulated.

\begin{figure}
    \centering
    \includegraphics[width=0.7\textwidth]{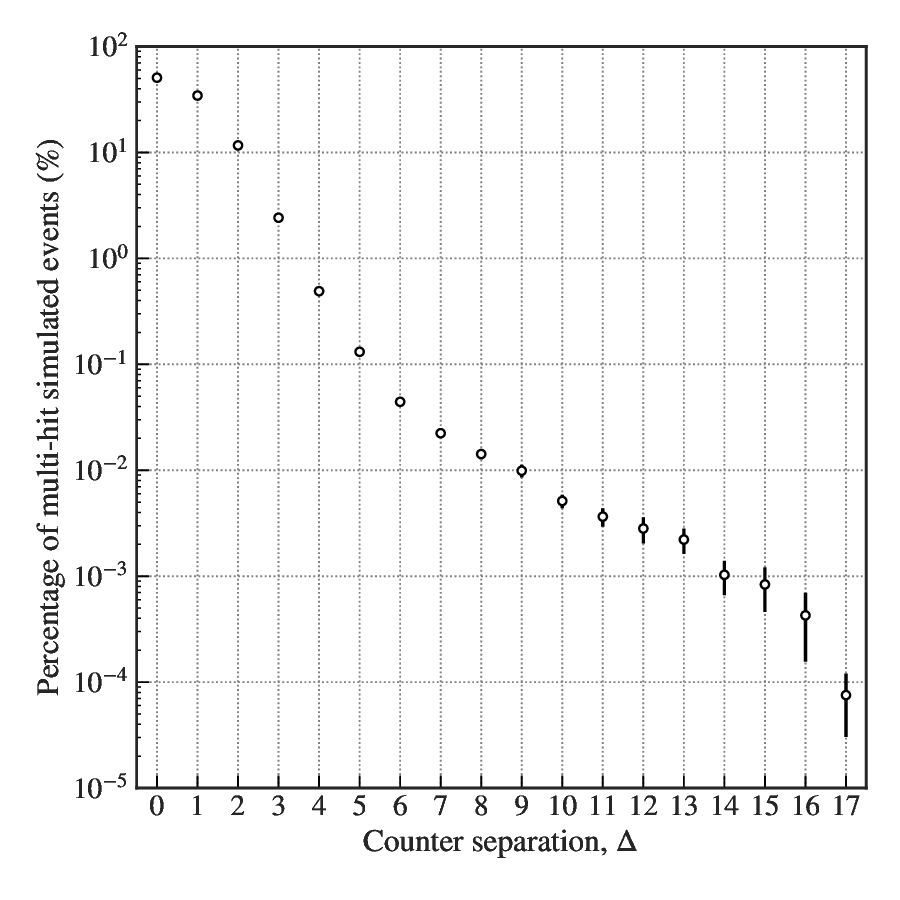}
    \caption{Percentage of  multi-hit simulated events as a function of the counter separation between hits. Error bars indicate the statistical uncertainties (one standard error of the mean). An event with multiple hits in multiple counters will contribute to several $\Delta$ percentages depending on the spatial distribution of the hits over the 18NM64.}
    \label{fig:MonteCarlo}
\end{figure}

To study the potential effect of the incident direction of the incoming secondary particles, neutrons were injected at the top of the detector with an incident angle from 0$^{\circ}$ (vertical incidence) to 45$^{\circ}$.
A power law spectrum of nucleons with a spectral index of 2.5 was assumed at the top of the detector
\citep{EvensonEAICRC21}.
Figure~\ref{fig:MonteCarlo} shows the dependence of the cross-multiplicity as a function of the counter separation $\Delta$ between hits. About half of the events with multiple hits have at least two hits in a single counter ($\Delta=0$), whereas $\approx 34\%$ of these events have hits in adjacent counters ($\Delta=1$). While single energetic secondary nucleons at $1<E<500$ GeV can produce events with long distance cross-multiplicity on rare occasion, their contribution is less than 0.1\% for $\Delta\ge6$ and is not sufficient to explain the observed deviation of $L_\Delta$ from 1 at large $\Delta$.

\subsection{Contributions of Single and Multiple Secondaries}

\begin{figure}
    \centering
    \includegraphics[width=\textwidth]{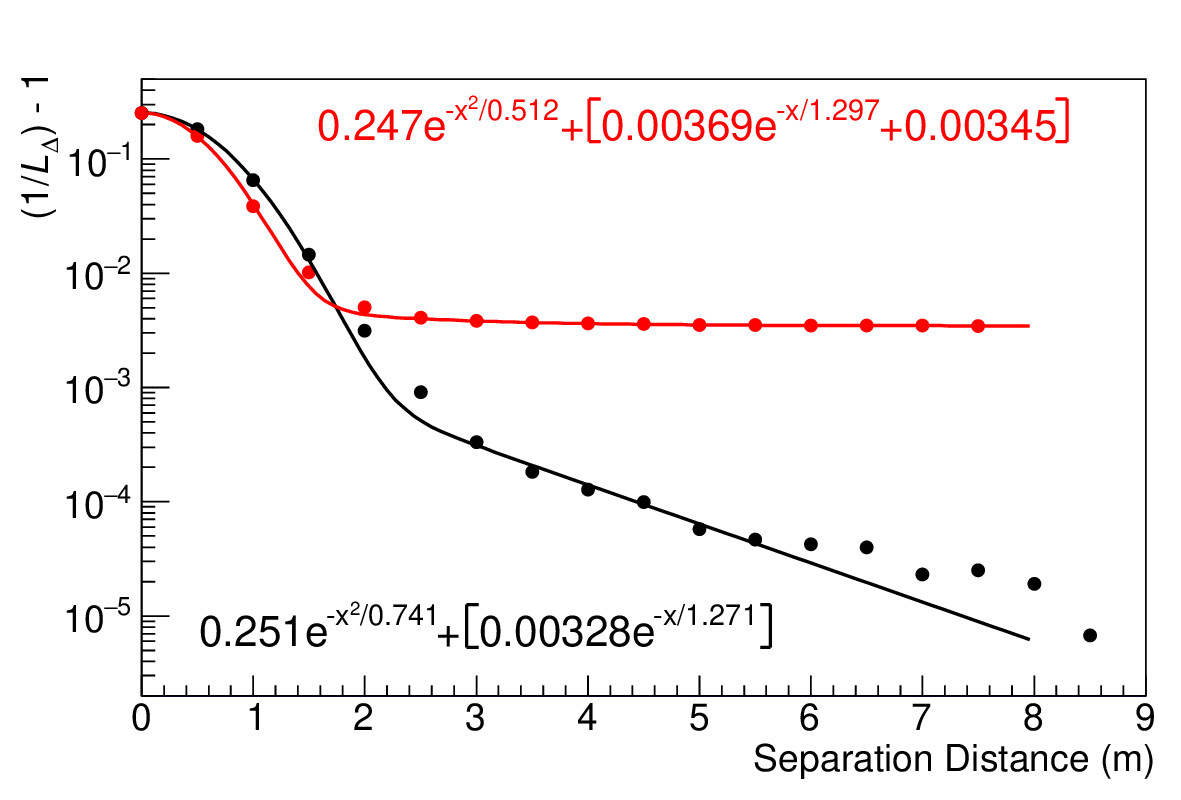}
    \caption{Detected followers per leader, $(1/L_\Delta)-1$, as a function of separation distance, $x$. 
    Note that $\Delta=x/{}$(50~cm).
    Measured data for $mm$ counter combinations (red circles) 
    are compared against results of MC simulations using single atmospheric secondary neutrons (black circles), scaled to the measured result at $x=0$.
    The first two terms of the fit to measurements (red curve and expression) are quantitatively similar to the fit to MC results (black curve and expression), so we attribute these terms to single secondaries from CR showers and the third term in the fit to measurements can be attributed to events with multiple secondaries.
    }
    \label{fig:MP}
\end{figure}

The average multiplicity is the inverse of the leader fraction, 1/$L_\Delta$. Figure~\ref{fig:MP} compares MC simulation results and measurements of the multiplicity minus one, which can also be expressed as $(1-L_\Delta)/L_\Delta$ and interpreted as the average number of followers per leader, as a function of separation distance (which is $\Delta$ times 0.5 m). 
These measurements, based on histograms such as shown in Figure~\ref{fig:xtd2}, 
are for $mm$ configurations, i.e., using middle tubes only, for all valid data 
between April 2018 and December 2024.
Here we scale the simulation results to match the measurement at separation distance $x= 0$. At small distance ($x\leq2$~m, i.e., $\Delta\leq4$), simulations and measurements agree reasonably well, while at larger distance the two are inconsistent. This discrepancy can be attributed to the effects of multiple associated atmospheric secondaries in the data, which are not included in the simulations as described above.

According to the simulation results shown in black in Figure~\ref{fig:MP}, we propose to empirically model the contribution from single secondaries with a sum of Gaussian and exponential terms.  
The former and the latter dominate the single-secondary contribution at smaller and larger separation distance, respectively. 
On the other hand, multiple secondaries associated with the same primary CRs should be distributed over a much larger length scale, and apparently their contribution to the far-counter multiplicity is well approximated by a constant, independent of separation distance. We therefore fit  $(1/L_{\Delta})-1$ with a function of the form $Qe^{-x^2/q}+Re^{-x/r}+S$, where $Q$, $q$, $R$, $r$, and $S$ are the fitting parameters and $x$ is expressed in m. Here the first two terms model the single-secondary contributions, while the constant term $S$ is for the multiple-secondary component, which is set to zero when we fit the simulation data. 

For single-step fitting, the small-separation data strongly dominate 
the fit, with unacceptable fitting results at large separation, so we solve this problem by using a two-step fitting procedure. First we fit $(1/L_{\Delta})-1$ with the sum of the exponential and constant terms at distances $x>3$~m, and then fix their parameters to the best-fit values before adding the Gaussian term and refitting over the entire range of separation distance. The best-fit functions are indicated in Figure~\ref{fig:MP}, in which the exponential and constant terms are shown in brackets to emphasize that they are fixed to the best-fit results at separation distance $x>3$~m. We obtain similar best-fit parameters for the Gaussian and exponential terms for measurements and simulations. The differences are within approximately 2\%, 30\%, 10\%, and 2\% for $Q$, $q$, $R$, and $r$, respectively, indicating good agreement regarding the contributions of single secondaries to  simulations and measured data.

The lateral distributions of secondary particles, especially neutrons, in the hadronic air showers have not been very well understood \citep{Schimassek2024,Pyras2025}. For the electromagnetic showers, a well known model describing the lateral distributions is the Nishimura-Kamata-Greisen (NKG) function~\citep{Greisen1960}, which contains a cusp around the shower core. The same functional form has been used for the approximation of the hadronic showers by combining parameters of many particle types, making the lateral distribution effectively wider depending on the energy of the primaries. Specifically, for secondary neutrons, the width is larger than and their arrival time is delayed compared to other particles because neutrons experience more scattering \citep{Schimassek2024}. 

In the present work, over the size of PSNM (9 m), we find the distribution of secondary nucleons from GeV-range cosmic ray showers to be independent of separation distance,
depending only on the position of the second counter (middle or end) as discussed in Section~\ref{sec:LvsDelta}.
From the far-counter leader fraction, we can estimate the probability that a given neutron count was followed by a count due to another atmospheric secondary particle {\it anywhere} in the 18NM64. 

As a first approximation, we estimate that probability by assuming that a different-secondary following count in any counter is an independent event.
In Section~\ref{sec:LvsDelta} we reported that 
$L_{\Delta\ge12}^{xe}\approx0.99706$ and $L_{\Delta\ge12}^{xm}\approx0.99654$. 
Each of these represents the probability that no different-secondary follower appeared on one of the 2 end or 16 middle counters, respectively, so the probability that no multiple-secondary follower appeared on {\it any} counter is estimated as $(L_{\Delta\ge12}^{xe})^2(L_{\Delta\ge12}^{xm})^{16} = 0.940$. 
This is the same regardless of the first counter ($x$ could be $m$ or $e$), so it applies to any count on any counter in the monitor.
Conversely, the probability that a given neutron count was associated with a later count in any counter from another secondary particle in the same shower is estimated as $P=0.060$.

That calculation assumes that each $L_{\Delta\geq12}$ value represents an uncorrelated probability.  
But after an ``original'' neutron count due to one secondary particle, sometimes another secondary particle accounts for later counts on {\it more than one} counter. 
To account for such ``double counting,'' a refined estimate of $P$ is
\begin{equation}
P \approx P_1+P_2+\dots+P_{18}-P_{1,2}-P_{2,1}-P_{2,3}-\dots-P_{17,16}-P_{17,18}-P_{18,17}.
\end{equation}
Here $P_i$ means the probability that the original neutron count (which could be on any counter) was followed by a later count on counter $i$ due to a different atmospheric secondary particle, and $P_{ij}$ is the double-counting probability of such a count on counter $i$ followed by yet another count on counter $j$.
The nearest neighbor terms $P_{i,i-1}+P_{i,i+1}$ effectively serve to correct $P_i$ for physically associated counts on any more distant counters, which were likely to be associated with a count on a nearest neighbor.  
Assuming the counts on $i$ and $i\pm1$ are from the same atmospheric secondary (but a different secondary from the one that produced the original count), we use $P_{i,i\pm1}/P_i \approx P_{\Delta=1}=1-L_{\Delta=1}$.
As noted earlier, $L_{\Delta=1}^{me}\approx L_{\Delta=1}^{mm}$, so we call this $L_{\Delta=1}^{mx}$.
We then obtain 
\begin{equation}
P \approx 2P_{\Delta\geq12}^{xe}\left(1-P_{\Delta=1}^{em}\right)
+16P_{\Delta\geq12}^{xm}\left(1-2P_{\Delta=1}^{mx}\right)=0.045,
\end{equation}
which expresses the summed probability of a later count on an end or middle counter due to a different atmospheric secondary from that producing the original count, with terms in parentheses to correct for double-counting.

In summary, the probability that a given count was associated with a later count anywhere in the 18NM64 from another secondary particle in the same shower is about 4.5\%.  This number is relevant for adjusting Monte Carlo predictions of the single-counter leader fraction, which has been used to study cosmic ray spectral variations, because the simulations so far only account for single atmospheric secondaries \citep{MangeardEA16b}.

\subsection{Time Series of Far-Counter Leader Fraction \label{sec:timeseries}}

\begin{figure}
    \centering
    \includegraphics[width=\textwidth]{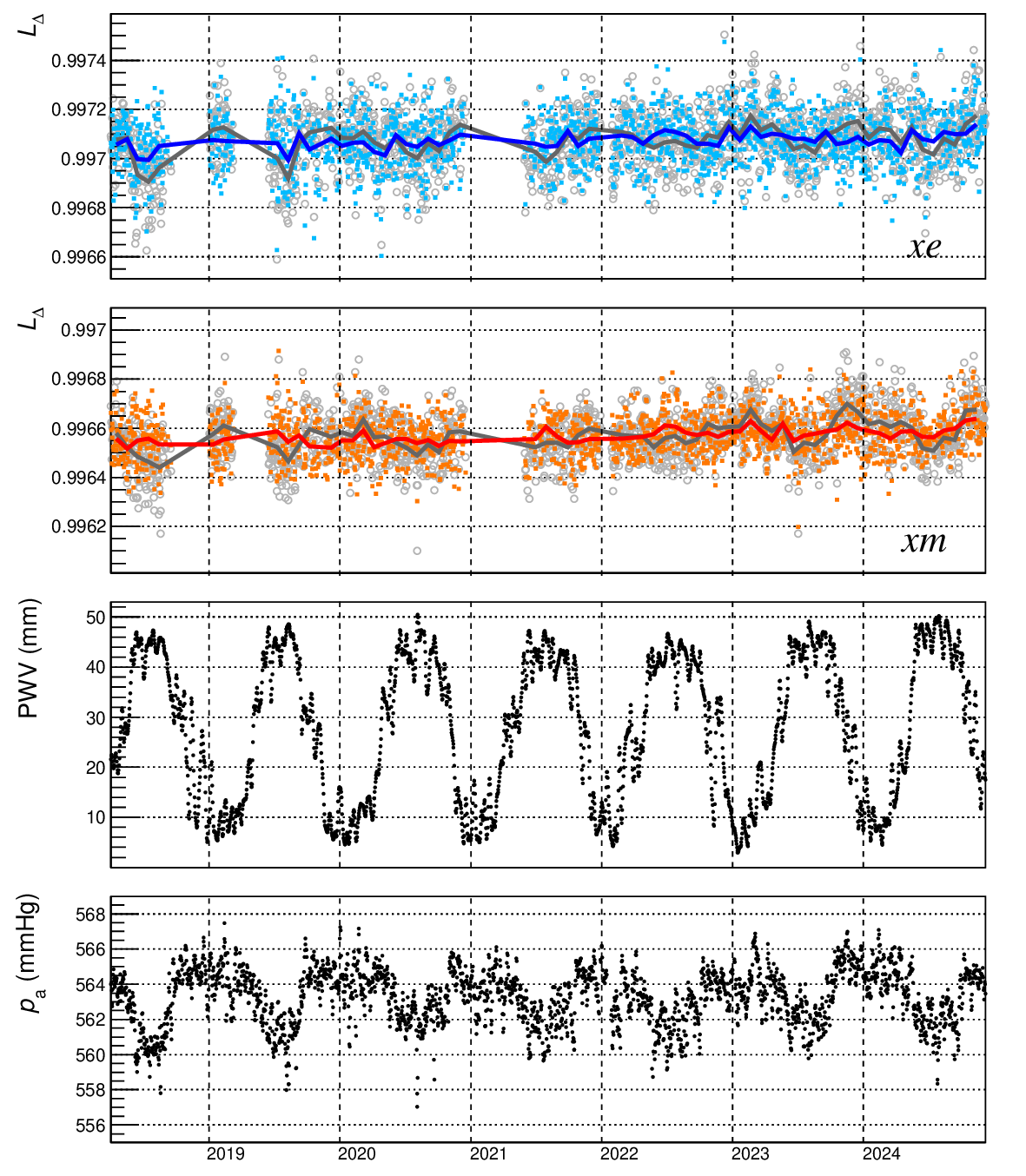}
    \caption{ (a) Time variation over 2018-2024 of the uncorrected (grey) and pressure-corrected (blue) far-counter leader fraction for the combinations ending with an end counter ($L_{\Delta\ge12}^{xe}$). (b) Similar to (a) but for the combinations ending with a middle counter ($L_{\Delta\ge12}^{xm}$) in red. For (a) and (b), the symbols are daily averages and the lines are 30-day averages. 
    After correction, the values show no obvious time dependence.
    (c) Triangular-smoothed precipitable water vapor, PWV, interpolated from the GDAS data but not used for correction, and (d) atmospheric pressure at PSNM, $p_\mathrm{a}$, used to perform data correction.}
    \label{fig:XLFvsTime}
\end{figure}

Figure~\ref{fig:XLFvsTime} shows the time series of the uncorrected and pressure-corrected far-counter ($\Delta\ge12$) leader fraction (see Equation~(\ref{eq:Lcorr})) for the $xe$ and $xm$ combinations, along with the precipitable water vapor (PWV) and atmospheric pressure ($p_\mathrm{a}$) in the two lower panels, which clearly vary annually due to seasons in Thailand. 
In Section~\ref{sec:corr} we described the correction procedure, and that a single correction for pressure variation effectively corrects for water vapor variations as well, because these variations are strongly correlated.
The daily averages of $L_{\Delta\ge12}$ have large statistical fluctuation and we do not observe the environmental effects. 
However, the 30-day averages of uncorrected rates (gray traces in Figure~\ref{fig:XLFvsTime}) are systematically higher at lower PWV and higher $p_\mathrm{a}$, and such dependence has been removed in the corrected rates (colored traces).

After the atmospheric corrections, the values of both $L_{\Delta\ge12}^{xe}$ and $L_{\Delta\ge12}^{xm}$ show no obvious time dependence over 2018-2024, varying from sunspot minimum to sunspot maximum conditions. 
To understand the non-observation of time dependence, we consider previous experience with the single-counter leader fraction $L$.  
In general, solar modulation, i.e., the effect of the solar activity cycle on the GCR flux and spectrum, decreases rapidly with increasing primary rigidity \citep[e.g.,][]{mangeard2018,poopakun2025}.
While there are strong time variations of $L$ in Antarctica, where the cutoff rigidity is $\sim$1~GV (an atmospheric cutoff, required for showers to produce neutron counts at ground level), such variations are much weaker for PSNM at Doi Inthanon, with the world's highest geomagnetic cutoff of $\sim$17 GV\@. 
Between sunspot minimum and maximum, the follower fraction $1-L$ at the South Pole NM was found to vary by roughly 0.006 out of 0.25, or about 2.4\%, while at PSNM it varies by roughly 0.0008 out of 0.2267, or about 0.35\% of $1-L$ \citep{BangliengApJ2020,Muangha2024}.  
Considering that the PSNM far-counter follower fraction is $\approx$0.003, variation of 0.35\% over the solar cycle would represent a far-counter follower variation $\sim10^{-5}$, which we are not able to measure.
Future work can explore whether time variation can be measured in the far-counter leader fraction at polar NM stations, which often have negligible water vapor effects and stronger physical GCR variations.  

\section{Conclusions}
Here we have reported a novel technique for a single NM station to analyze temporal and spatial association between consecutive events detected by different neutron counters.  
We report the cross-counter leader fraction, which measures the inverse multiplicity of recording a count first on one counter and then on another, due to the same CR shower. 
At PSNM, associated neutrons within $<$2~m are mainly the tertiary products of a single secondary CR interacting with the lead producer, while associated neutrons at $>$3~m are dominated by multiple secondary particles from the same primary CR shower. 
The contribution to the observed spatial distribution of neutrons from a single secondary particle is consistent with that from our Monte Carlo simulations, and can be represented as a Gaussian plus exponential function of counter separation distance. 
For multiple secondaries, we infer a contribution that is independent of counter separation distance, with some delay in the timing distribution of associated neutron counts, which can be attributed to a fraction of later secondary particles that are slow (non-relativistic) neutrons.
A fraction of 0.003 of overall neutron counts are associated with a following count on each individual far counter, and for the 18-counter PSNM, we estimate that 4.5\% of all neutron counts are associated with a later count from another secondary particle on at least one of the counters.
We have demonstrated an upgrade of NM capabilities from improved electronics with absolute timing, and no other hardware costs, which could be installed at other NMs worldwide.
Further experiments and modeling will be performed to better understand the behavior of NMs and CR showers and to improve our Monte Carlo simulations, including accounting for multiple secondary particles and associated multiple counts and time delays in the NM.

\begin{acknowledgments}
This research was supported in Thailand by the National Science and Technology Development Agency (NSTDA) and the National Research Council of Thailand (NRCT) under the High-Potential Research Team Grant Program (N42A650868) and
from the NSRF via the Research and Innovation Acceleration Agency for Competitiveness and Area Development (RCAD) (Program Management Unit for Technology and Innovation for Future Industries (PMU-B): Brainpower for Future Industries) [grant number B39G690003], and by the US NSF Award for Collaborative Research: The Simpson Neutron
Monitor Network (2112439).
\end{acknowledgments}

\bibliography{xlf}{}

@ARTICLE{AhdidaEA2022,
AUTHOR={Ahdida, C. and Bozzato, D. and Calzolari, D. and Cerutti, F. and Charitonidis, N. and Cimmino, A. and Coronetti, A. and D’Alessandro, G. L. and Donadon Servelle, A. and Esposito, L. S. and Froeschl, R. and García Alía, R. and Gerbershagen, A. and Gilardoni, S. and Horváth, D. and Hugo, G. and Infantino, A. and Kouskoura, V. and Lechner, A. and Lefebvre, B. and Lerner, G. and Magistris, M. and Manousos, A. and Moryc, G. and Ogallar Ruiz, F. and Pozzi, F. and Prelipcean, D. and Roesler, S. and Rossi, R. and Sabaté Gilarte, M. and Salvat Pujol, F. and Schoofs, P. and Stránský, V. and Theis, C. and Tsinganis, A. and Versaci, R. and Vlachoudis, V. and Waets, A. and Widorski, M.},    
TITLE={New Capabilities of the {FLUKA} Multi-Purpose Code},      
JOURNAL={Front.\ Phys.},      
VOLUME={9},           
YEAR={2022},      
cURL={https://www.frontiersin.org/articles/10.3389/fphy.2021.788253},       
DOI={10.3389/fphy.2021.788253},      
ISSN={2296-424X},  
}

@ARTICLE{Aiemsa-adEA15,
       author = {{Aiemsa-ad}, N. and {Ruffolo}, D. and {S{\'a}iz}, A. and {Mangeard}, P. -S. and {Nutaro}, T. and {Nuntiyakul}, W. and {Kamyan}, N. and {Khumlumlert}, T. and {Kr{\"u}ger}, H. and {Moraal}, H. and {Bieber}, J.~W. and {Clem}, J. and {Evenson}, P.},
        title = {Measurement and simulation of neutron monitor count rate dependence on surrounding structure},
      journal = {J.~Geophys.\ Res.\ (Space Phys.)},
         year = 2015,
        nomonth = jul,
       volume = {120},
       number = {7},
        pages = {5253--5265},
          doi = {10.1002/2015JA021249},
       adsurl = {https://ui.adsabs.harvard.edu/abs/2015JGRA..120.5253A}
}

@mastersthesis{Amratisha2025,
       author = {{Amratisha}, Koth},
        title = "{}",
       school = {Mahidol University},
         year = 2025,
        month = dec
}

@INPROCEEDINGS{Balabin2008,
       author = {{Balabin}, Yu. V. and {Gvozdevsky}, B.~B. and {Vashenyuk}, E.~V. and {Schur}, L.~I.},
        title = "{Observing Multiplicity Effect during December 13, 2006 event on the Barentsburg Neutron Monitor}",
    booktitle = {Proceedings of the 30th International Cosmic Ray Conference},
         year = 2008,
       seriesbak = {International Cosmic Ray Conference},
       volume = {1},
        month = jan,
        pages = {257-260},
       adsurl = {https://ui.adsabs.harvard.edu/abs/2008ICRC....1..257B}
}

@ARTICLE{BangliengApJ2020,
       author = {{Banglieng}, C. and {Janthaloet}, H. and {Ruffolo}, D. and {S{\'a}iz}, A. and {Mitthumsiri}, W. and {Muangha}, P. and {Evenson}, P. and {Nutaro}, T. and {Pyle}, R. and {Seunarine}, S. and {Madsen}, J. and {Mangeard}, P. -S. and {Macatangay}, R.},
        title = {Tracking Cosmic-Ray Spectral Variation during 2007--2018 Using Neutron Monitor Time-delay Measurements},
      journal = {\apj},
         year = 2020,
        nomonth = feb,
       volume = {890},
       number = {1},
          eid = {21},
        pages = {21},
          doi = {10.3847/1538-4357/ab6661},
       adsurl = {https://ui.adsabs.harvard.edu/abs/2020ApJ...890...21B}
}

@ARTICLE{BATTISTONI201510,
title = {Overview of the {FLUKA} code},
journal = {Ann.\ Nucl.\ Energy},
volume = {82},
pages = {10--18},
year = {2015},
note = {Joint International Conference on Supercomputing in Nuclear Applications and Monte Carlo 2013, SNA + MC 2013. Pluri- and Trans-disciplinarity, Towards New Modeling and Numerical Simulation Paradigms},
issn = {0306-4549},
doi = {10.1016/j.anucene.2014.11.007},
author = {Giuseppe Battistoni and Till Boehlen and Francesco Cerutti and Pik Wai Chin and Luigi Salvatore Esposito and Alberto Fassò and Alfredo Ferrari and Anton Lechner and Anton Empl and Andrea Mairani and Alessio Mereghetti and Pablo Garcia Ortega and Johannes Ranft and Stefan Roesler and Paola R. Sala and Vasilis Vlachoudis and George Smirnov},
}

@ARTICLE{Bieber2004,
       author = {{Bieber}, J.~W. and {Clem}, J.~M. and {Duldig}, M.~L. and {Evenson}, P.~A. and {Humble}, J.~E. and {Pyle}, R.},
        title = "{Latitude survey observations of neutron monitor multiplicity}",
      journal = {\jgr\ (Space Phys.)},
         year = 2004,
        month = dec,
       volume = {109},
       number = {A12},
          eid = {A12106},
        pages = {A12106},
          doi = {10.1029/2004JA010493},
       adsurl = {https://ui.adsabs.harvard.edu/abs/2004JGRA..10912106B}
}

@article{Chaiwongkhot2021,
   author = {K. Chaiwongkhot and David Ruffolo and W. Yamwong and J. Prabket and Pierre-Simon Mangeard and Alejandro Sáiz and Warit Mitthumsiri and Chanoknan Banglieng and E. Kittiya and Waraporn Nuntiyakul and U. Tippawan and M. Jitpukdee and S. Aukkaravittayapun},
   doi = {10.1016/j.astropartphys.2021.102617},
   issn = {09276505},
   issue = {June},
   journal = {Astropart.\ Phys.},
   pages = {102617},
   publisher = {Elsevier B.V.},
   title = {Measurement and simulation of the neutron propagation time distribution inside a neutron monitor},
   volume = {132},
   year = {2021},
}

@article{Chaiwongkhot2024,
  author = "Chaiwongkhot, Kullapha  and  Ruffolo, David  and  S\'aiz, Alejandro  and  Mitthumsiri, Warit  and  Chaiwongkhot, Poompong  and  Banglieng, Chanoknan  and  Mangeard, Pierre-Simon  and  Evenson, Paul  and  Lakronwat, Jidapa",
  title = "{Measurement of sparse vs. dense atmospheric secondary particles from cosmic ray showers using coincident signals on various counters in a neutron monitor}",
  doi = "10.22323/1.444.1264",
  journal = "PoS",
  year = 2023,
  volume = "ICRC2023",
  pages = "1264"
}

@ARTICLE{Clem2000,
       author = {{Clem}, John M. and {Dorman}, Lev I.},
        title = "{Neutron Monitor Response Functions}",
      journal = {\ssr},
         year = 2000,
        month = jul,
       volume = {93},
        pages = {335-359},
          doi = {10.1023/A:1026508915269},
       adsurl = {https://ui.adsabs.harvard.edu/abs/2000SSRv...93..335C}
}

@ARTICLE{Crosby2024,
       author = {{Crosby}, N. and {Mavromichalaki}, H. and {Malandraki}, O. and {Gerontidou}, M. and {Karavolos}, M. and {Lingri}, D. and {Makrantoni}, P. and {Papailiou}, M. and {Paschalis}, P. and {Tezari}, A.},
        title = "{Very High Energy Solar Energetic Particle Events and Ground Level Enhancement Events: Forecasting and Alerts}",
      journal = {Space Weather},
         year = 2024,
        month = sep,
       volume = {22},
       number = {9},
          eid = {e2023SW003839},
        pages = {e2023SW003839},
          doi = {10.1029/2023SW003839},
       adsurl = {https://ui.adsabs.harvard.edu/abs/2024SpWea..2203839C}
}

@article{EvensonEAICRC21,
  author = "Evenson, Paul  and  Clem, John  and  Mangeard, Pierre-Simon  and  Nuntiyakul, Waraporn  and  Ruffolo, David  and  S\'aiz, Alejandro  and  Seripienlert, Achara  and  Seunarine, Surujhdeo",
  title = "{Multiple Particle Detection in a Neutron Monitor}",
  doi = "10.22323/1.395.1240",
  journal = "PoS",
  year = 2021,
  volume = "ICRC2021",
  pages = "1240"
}

@article{Greisen1960,
   author = "Greisen, K",
   title = "Cosmic Ray Showers", 
   journal= "Annu.\ Rev.\ Nucl.\ Part.\ Sci.",
   year = "1960",
   volume = "10",
   number = "Volume 10, ",
   pages = "63-108",
   doi = "10.1146/annurev.ns.10.120160.000431",
   url = "https://www.annualreviews.org/content/journals/10.1146/annurev.ns.10.120160.000431",
   publisher = "Annual Reviews",
   issn = "1545-4134",
   type = "Journal Article",
  }

@ARTICLE{Hayashi2026,
       author = {{Hayashi}, Y. and {Munakata}, K. and {Kozai}, M. and {Kataoka}, R. and {Kadokura}, A. and {Kato}, C. and {Miyashita}, N. and {Miyake}, S. and {Murase}, K. and {Duldig}, M.~L. and {Ruffolo}, D. and {Mitthumsiri}, W. and {Muangha}, P. and {S{\'a}iz}, A. and {Seunarine}, S. and {Evenson}, P. and {Mangeard}, P.-S. and {Iwai}, K. and {Menjo}, H. and {Echer}, E. and {Dal Lago}, A. and {Rockenbach}, M. and {Schuch}, N.~J. and {Bageston}, J.~V. and {Braga}, C.~R. and {Al Jassar}, H.~K. and {Sharma}, M.~M. and {Burahmah}, N. and {Zaman}, F. and {Sabbah}, I. and {Kuwabara}, T. and {Chen}, D. and {Huang}, J.},
        title = "{Real-time monitoring of the rigidity spectrum of large Forbush decreases in May and October 2024 with the paired neutron monitor and muon detector at the Antarctic Syowa Station}",
      journal = {Earth, Planets and Space},
     xkeywords = {Forbush effect, Cosmic rays, Space weather, Cosmic ray detectors, Solar-terrestrial interactions, 546, 329, 2037, 325, 1473, High Energy Astrophysical Phenomena},
         year = 2026,
        xmonth = jun,
       xvolume = {986},
       xnumber = {1},
          xeid = {L7},
        xpages = {L7},
          doi = {10.1186/s40623-026-02386-y},
         note = {In press}
}

@ARTICLE{Khamphakdee2025,
       author = {{Khamphakdee}, S. and {Nuntiyakul}, W. and {Banglieng}, C. and {Seripienlert}, A. and {Yakum}, P. and {S{\'a}iz}, A. and {Ruffolo}, D. and {Evenson}, P. and {Munakata}, K. and {Komonjinda}, S.},
        title = "{Methods for Tracking Cosmic-Ray Spectral Changes Using Neutron Monitors at High Cutoff Rigidity}",
      journal = {\apj},
         year = 2025,
        month = may,
       volume = {984},
       number = {1},
          eid = {51},
        pages = {51},
          doi = {10.3847/1538-4357/adc5f9},
       adsurl = {https://ui.adsabs.harvard.edu/abs/2025ApJ...984...51K}
}

@ARTICLE{Kollar2011,
       author = {{Koll{\'a}r}, V. and {Kudela}, K. and {Minarovjech}, M.},
        title = "{Some alternative instrumentation for galactic cosmic rays measurement using ground based neutron monitor detectors. I. Elapsed time methods}",
      journal = {Contrib.\ Astron.\ Obs.\ Skaln.\ Pleso},
         year = 2011,
        month = may,
       volume = {41},
       number = {1},
        pages = {5-14},
       adsurl = {https://ui.adsabs.harvard.edu/abs/2011CoSka..41....5K}
}

@article{Kuwabara2006a,
author = {Kuwabara, T. and Bieber, J. W. and Clem, J. and Evenson, P. and Pyle, R. and Munakata, K. and Yasue, S. and Kato, C. and Akahane, S. and Koyama, M. and Fujii, Z. and Duldig, M. L. and Humble, J. E. and Silva, M. R. and Trivedi, N. B. and Gonzalez, W. D. and Schuch, N. J.},
title = {Real-time cosmic ray monitoring system for space weather},
journal = {Space Weather},
volume = {4},
number = {8},
eid = {S08001},
pages = {S08001},
doi = {10.1029/2005SW000204},
url = {https://agupubs.onlinelibrary.wiley.com/doi/abs/10.1029/2005SW000204},
eprint = {https://agupubs.onlinelibrary.wiley.com/doi/pdf/10.1029/2005SW000204},
year = {2006}
}

@article{Kuwabara2006b,
author = {Kuwabara, T. and Bieber, J. W. and Clem, J. and Evenson, P. and Pyle, R.},
title = {Development of a ground level enhancement alarm system based upon neutron monitors},
journal = {Space Weather},
volume = {4},
number = {10},
eid = {S10001},
pages = {S10001},
doi = {10.1029/2006SW000223},
url = {https://agupubs.onlinelibrary.wiley.com/doi/abs/10.1029/2006SW000223},
eprint = {https://agupubs.onlinelibrary.wiley.com/doi/pdf/10.1029/2006SW000223},
year = {2006}
}

@ARTICLE{Linsley1984,
       author = {{Linsley}, J.},
        title = "{Sub-luminal pulses from cosmic-ray air showers}",
      journal = {J.~Phys.~G: Nucl.\ Phys.},
         year = 1984,
        month = aug,
       volume = {10},
       number = {8},
        pages = {L191-L195},
          doi = {10.1088/0305-4616/10/8/005},
       adsurl = {https://ui.adsabs.harvard.edu/abs/1984JPhG...10L.191L}
}

@ARTICLE{Mangeard2016,
       author = {{Mangeard}, P. -S. and {Ruffolo}, D. and {S{\'a}iz}, A. and {Madlee}, S. and {Nutaro}, T.},
        title = {Monte {C}arlo simulation of the neutron monitor yield function},
      journal = {J.~Geophys.\ Res.\ (Space Phys.)},
         year = 2016,
        nomonth = aug,
       volume = {121},
       number = {8},
        pages = {7435--7448},
          doi = {10.1002/2016JA022638},
       adsurl = {https://ui.adsabs.harvard.edu/abs/2016JGRA..121.7435M}
}

@ARTICLE{MangeardEA16b,
       author = {{Mangeard}, P. -S. and {Ruffolo}, D. and {S{\'a}iz}, A. and {Nuntiyakul}, W. and {Bieber}, J.~W. and {Clem}, J. and {Evenson}, P. and {Pyle}, R. and {Duldig}, M.~L. and {Humble}, J.~E.},
        title = {Dependence of the neutron monitor count rate and time delay distribution on the rigidity spectrum of primary cosmic rays},
      journal = {J.~Geophys.\ Res.\ (Space Phys.)},
         year = 2016,
        nomonth = dec,
       volume = {121},
       number = {12},
        pages = {11,620--11,636},
          doi = {10.1002/2016JA023515},
       adsurl = {https://ui.adsabs.harvard.edu/abs/2016JGRA..12111620M}
}

@ARTICLE{Mangeard2018,
       author = {{Mangeard}, Pierre-Simon and {Clem}, John and {Evenson}, Paul and {Pyle}, Roger and {Mitthumsiri}, Warit and {Ruffolo}, David and {S{\'a}iz}, Alejandro and {Nutaro}, Tanin},
        title = "{Distinct Pattern of Solar Modulation of Galactic Cosmic Rays above a High Geomagnetic Cutoff Rigidity}",
      journal = {\apj},
         year = 2018,
        month = may,
       volume = {858},
       number = {1},
          eid = {43},
        pages = {43},
          doi = {10.3847/1538-4357/aabd3c},
       adsurl = {https://ui.adsabs.harvard.edu/abs/2018ApJ...858...43M}
}

@ARTICLE{Mishev2013,
       author = {{Mishev}, A.~L. and {Usoskin}, I.~G. and {Kovaltsov}, G.~A.},
        title = "{Neutron monitor yield function: New improved computations}",
      journal = {\jgr\ (Space Phys.)},
         year = 2013,
        month = jun,
       volume = {118},
       number = {6},
        pages = {2783-2788},
          doi = {10.1002/jgra.50325},
archivePrefix = {arXiv},
       eprint = {1612.01498},
 primaryClass = {physics.space-ph},
       adsurl = {https://ui.adsabs.harvard.edu/abs/2013JGRA..118.2783M}
}

@ARTICLE{Mishev2024,
       author = {{Mishev}, Alexander and {Larsen}, Nicholas and {Asvestari}, Eleanna and {S{\'a}iz}, Alejandro and {Ann Shea}, Margaret and {Strauss}, Du Toit and {Ruffolo}, David and {Banglieng}, Chanoknan and {Seunarine}, Surujhdeo and {Duldig}, Marc L. and {Gil}, Agnieszka and {Blanco}, Juan Jos{\'e} and {Garc{\'\i}a-Poblaci{\'o}n}, Oscar and {Cervino-Solana}, Pablo and {Adams}, Jr., James H. and {Usoskin}, Ilya},
        title = "{Anisotropic Forbush decrease of 24 March 2024: First look}",
      journal = {Adv.\ Space Res.},
         year = 2024,
        month = oct,
       volume = {74},
       number = {8},
        pages = {4160-4172},
          doi = {10.1016/j.asr.2024.08.027},
       adsurl = {https://ui.adsabs.harvard.edu/abs/2024AdSpR..74.4160M}
}

@ARTICLE{Mitthumsiri2025,
       author = {{Mitthumsiri}, W. and {Ruffolo}, D. and {Munakata}, K. and {Kozai}, M. and {Hayashi}, Y. and {Kato}, C. and {Muangha}, P. and {S{\'a}iz}, A. and {Evenson}, P. and {Mangeard}, P.-S. and {Clem}, J. and {Seunarine}, S. and {Nuntiyakul}, W. and {Miyashita}, N. and {Kataoka}, R. and {Kadokura}, A. and {Miyake}, S. and {Iwai}, K. and {Menjo}, H. and {Echer}, E. and {Dal Lago}, A. and {Rockenbach}, M. and {Schuch}, N.~J. and {Bageston}, J.~V. and {Braga}, C.~R. and {Al Jassar}, H.~K. and {Sharma}, M.~M. and {Burahmah}, N. and {Zaman}, F. and {Duldig}, M.~L. and {Sabbah}, I. and {Kuwabara}, T.},
        title = "{Ground-based Observations of Temporal Variation of the Cosmic-Ray Spectrum during Forbush Decreases}",
      journal = {\apjl},
         year = 2025,
        month = jun,
       volume = {986},
       number = {1},
          eid = {L7},
        pages = {L7},
          doi = {10.3847/2041-8213/add7d1},
archivePrefix = {arXiv},
       eprint = {2505.08248},
 primaryClass = {astro-ph.HE},
       adsurl = {https://ui.adsabs.harvard.edu/abs/2025ApJ...986L...7M}
}

@ARTICLE{Moraal2000,
       author = {{Moraal}, H. and {Belov}, A. and {Clem}, J.~M.},
        title = "{Design and co-Ordination of Multi-Station International Neutron Monitor Networks}",
      journal = {\ssr},
         year = 2000,
        month = jul,
       volume = {93},
        pages = {285-303},
          doi = {10.1023/A:1026504814360},
       adsurl = {https://ui.adsabs.harvard.edu/abs/2000SSRv...93..285M}
}

@ARTICLE{Muangha2024,
       author = {{Muangha}, Pradiphat and {Ruffolo}, David and {S{\'a}iz}, Alejandro and {Banglieng}, Chanoknan and {Evenson}, Paul and {Seunarine}, Surujhdeo and {Oh}, Suyeon and {Jung}, Jongil and {Duldig}, Marc L. and {Humble}, John E.},
        title = "{Variations in the Inferred Cosmic-Ray Spectral Index as Measured by Neutron Monitors in Antarctica}",
      journal = {\apj},
         year = 2024,
        month = oct,
       volume = {974},
       number = {2},
          eid = {284},
        pages = {284},
          doi = {10.3847/1538-4357/ad73d6},
archivePrefix = {arXiv},
       eprint = {2408.13999},
 primaryClass = {astro-ph.HE},
       adsurl = {https://ui.adsabs.harvard.edu/abs/2024ApJ...974..284M}
}

@ARTICLE{Poopakun2025,
       author = {{Poopakun}, K. and {Nuntiyakul}, W. and {Kato}, C. and {Munakata}, K. and {Kozai}, M. and {Hayashi}, Y. and {Ruffolo}, D. and {Iwai}, K. and {Menjo}, H. and {Zhai}, L.~M. and {Duldig}, M.~L.},
        title = "{Solar and Interplanetary Determinants of Long-term Solar Modulation of Cosmic-Ray Intensity for Median Rigidities of 11{\textendash}107 GV}",
      journal = {\apj},
         year = 2025,
        month = oct,
       volume = {991},
       number = {2},
          eid = {127},
        pages = {127},
          doi = {10.3847/1538-4357/adfdd3},
       adsurl = {https://ui.adsabs.harvard.edu/abs/2025ApJ...991..127P}
}

@article{Pyras2025,
  author = "Pyras, Lilly  and  Soldin, Dennis  and  Riehn, Felix",
  title = "{Probing hadronic interactions with high-energy and low-energy muons in extensive air showers}",
  doi = "10.22323/1.501.0368",
  journal = "PoS",
  year = 2025,
  volume = "ICRC2025",
  pages = "368"
}

@article{TimeDelay,
	author = {David {Ruffolo} and Alejandro {S{\'a}iz} and Pierre-Simon {Mangeard} and Nattapong {Kamyan} and Pradiphat {Muangha} and Tanin {Nutaro} and Supon {Sumran} and Chaweewan {Chaiwattana} and Nipon {Gasiprong} and Changrueangrit {Channok} and Cherdchai {Wuttiya} and Manit {Rujiwarodom} and Paisan {Tooprakai} and Burin {Asavapibhop} and John W. {Bieber} and John {Clem} and Paul {Evenson} and Kazuoki {Munakata}},
	doi = {10.3847/0004-637X/817/1/38},
	eid = {38},
	issn = {1538-4357},
	journal = {\apj},
	nomonth = jan,
	number = {1},
	publisher = {IOP Publishing},
	title = {Monitoring short-term cosmic-ray spectral variations using neutron monitor time-delay measurements},
	volume = 817,
    pages = 38,
	year = {2016}
}

@article{Saiz2017,
  author = "S\'aiz, Alejandro  and  Mitthumsiri, Warit  and  Ruffolo, David  and  Evenson, Paul  and  Nutaro, Tanin",
  title = "{Measurement of cross-counter leader fractions in an 18NM64: Detecting single and multiple atmospheric secondaries}",
  doi = "10.22323/1.301.0047",
  journal = "PoS",
  year = 2017,
  volume = "ICRC2017",
  pages = "047"
}

@ARTICLE{Schimassek2024,
       author = {{Schimassek}, Martin and {Engel}, Ralph and {Ferrari}, Alfredo and {Roth}, Markus and {Schmidt}, David and {Veberi{\v{c}}}, Darko},
        title = "{Neutron Production in Simulations of Extensive Air Showers}",
      journal = {arXiv e-prints},
         year = 2024,
        month = jun,
          eid = {arXiv:2406.11702},
        pages = {arXiv:2406.11702},
          doi = {10.48550/arXiv.2406.11702},
archivePrefix = {arXiv},
       eprint = {2406.11702},
 primaryClass = {hep-ph},
       adsurl = {https://ui.adsabs.harvard.edu/abs/2024arXiv240611702S}
}

@ARTICLE{Shea2012,
       author = {{Shea}, M.~A. and {Smart}, D.~F.},
        title = "{Space Weather and the Ground-Level Solar Proton Events of the 23rd Solar Cycle}",
      journal = {\ssr},
         year = 2012,
        month = oct,
       volume = {171},
       number = {1-4},
        pages = {161-188},
          doi = {10.1007/s11214-012-9923-z},
       adsurl = {https://ui.adsabs.harvard.edu/abs/2012SSRv..171..161S}
}

@ARTICLE{Simpson2000,
       author = {{Simpson}, John A.},
        title = "{The Cosmic Ray Nucleonic Component: The Invention and Scientific Uses of the Neutron Monitor - (Keynote Lecture)}",
      journal = {\ssr},
         year = 2000,
        month = jul,
       volume = {93},
        pages = {11-32},
          doi = {10.1023/A:1026567706183},
       adsurl = {https://ui.adsabs.harvard.edu/abs/2000SSRv...93...11S}
}

@ARTICLE{Smart2000,
       author = {{Smart}, D.~F. and {Shea}, M.~A. and {Fl{\"u}ckiger}, E.~O.},
        title = "{Magnetospheric Models and Trajectory Computations}",
      journal = {\ssr},
         year = 2000,
        month = jul,
       volume = {93},
        pages = {305-333},
          doi = {10.1023/A:1026556831199},
       adsurl = {https://ui.adsabs.harvard.edu/abs/2000SSRv...93..305S}
}

@ARTICLE{Strauss2020,
       author = {{Strauss}, Du Toit and {Poluianov}, Stepan and {van der Merwe}, Cobus and {Kr{\"u}ger}, Hendrik and {Diedericks}, Corrie and {Kr{\"u}ger}, Helena and {Usoskin}, Ilya and {Heber}, Bernd and {Nndanganeni}, Rendani and {Blanco-{\'A}valos}, Juanjo and {Garc{\'\i}a-Tejedor}, Ignacio and {Herbst}, Konstantin and {Caballero-Lopez}, Rogelio and {Moloto}, Katlego and {Lara}, Alejandro and {Walter}, Michael and {Giday}, Nigussie Mezgebe and {Traversi}, Rita},
        title = "{The mini-neutron monitor: a new approach in neutron monitor design}",
      journal = {J.~Space Weather Space Clim.},
         year = 2020,
        month = jul,
       volume = {10},
          eid = {39},
        pages = {39},
          doi = {10.1051/swsc/2020038},
       adsurl = {https://ui.adsabs.harvard.edu/abs/2020JSWSC..10...39S}
}

@ARTICLE{Tongiorgi1949,
       author = {{Tongiorgi}, Vanna Cocconi},
        title = "{Neutrons in the Extensive Air Showers of the Cosmic Radiation}",
      journal = {Phys.\ Rev.},
         year = 1949,
        month = may,
       volume = {75},
       number = {10},
        pages = {1532-1540},
          doi = {10.1103/PhysRev.75.1532},
       adsurl = {https://ui.adsabs.harvard.edu/abs/1949PhRv...75.1532T}
}
\bibliographystyle{aasjournalv7}

\end{document}